\documentclass[amsmath,amssymb,aps,prl,twocolumn]{revtex4-1}

\ifx\pdfoutput\undefined
\usepackage[dvipdfmx]{graphicx}
\usepackage[dvipdfmx]{hyperref}
\else
\usepackage{graphicx}
\usepackage{hyperref}
\fi

\usepackage{txfonts}
\usepackage[T1]{fontenc}
\usepackage{xspace}

\usepackage{dcolumn}
\usepackage{bm}
\usepackage{amsmath}
\usepackage{booktabs}
\usepackage[usenames]{color}
\usepackage{xcolor}

\newcommand{\nn}{\nonumber \\}
\renewcommand{\v}[1]{{\bf #1}}

\begin{document}


\title{Spin Thermal Hall Conductivity of a Kagom\'e Antiferromagnet}

\author{Hayato Doki$^1$}
\author{Masatoshi Akazawa$^1$}
\author{Hyun-Yong Lee$^1$}
\author{Jung Hoon Han$^2$}
\author{Kaori Sugii$^1$}
\author{Masaaki Shimozawa$^1$}
\author{Naoki Kawashima$^1$}
\author{Migaku Oda$^3$}
\author{Hiroyuki Yoshida$^3$}
\author{Minoru Yamashita$^1$}
\email[]{my@issp.u-tokyo.ac.jp}

\affiliation{
$^1$The Institute for Solid State Physics, The University of Tokyo, Kashiwa, 277-8581, Japan\\
$^2$Department of Physics, Sungkyunkwan University, Suwon 16419, Korea\\
$^3$Department of Physics, Faculty of Science, Hokkaido University, Sapporo 060-0810, Japan
}

\date{\today}

\begin{abstract}
A clear thermal Hall signal ($\kappa_{xy}$) was observed in the spin liquid phase of the $S=1/2$ kagom\'e antiferromagnet Ca kapellasite (CaCu$_3$(OH)$_6$Cl$_2\cdot 0.6$H$_2$O).
We found that $\kappa_{xy}$ is well reproduced, both qualitatively and quantitatively, using the Schwinger-boson mean-field theory with the Dzyaloshinskii--Moriya interaction of $D/J \sim 0.1$.
In particular, $\kappa_{xy}$ values of Ca kapellasite and those of another kagom\'e antiferromagnet, volborthite, converge to one single curve in simulations modeled using Schwinger bosons, indicating a common temperature dependence of $\kappa_{xy}$ for the spins of a kagom\'e antiferromagnet.
\end{abstract}

\pacs{}
\maketitle

Identifying the ground state of a kagom\'e Heisenberg antiferromagnet (KHA) has been a central issue of condensed-matter physics because the KHA is expected to host many unknown spin-liquid phases, including the resonating-valence-bond state \cite{Jiang2012}, Z$_2$ spin liquids \cite{Sachdev1992, WangVishwanath2006, YanHuseWhite2011}, chiral spin liquids \cite{YinChenHe2014, Bauer2014, Messio2017}, fermionic spin liquids \cite{LiErnChern2017}, and Dirac spin liquids \cite{Jiang2016, Liao2017, Iqbal2013, WeiZhu2018,YinChenHe2017}.
It is therefore very important to find the elementary excitations expected to emerge in these unknown phases through experiments on ideal kagom\'e materials.
So far, herbertsmithite ZnCu$_3$(OH)$_6$Cl$_2$ \cite{Shores2005} is the best-studied compound in which the spins remain disordered down to the lowest temperature \cite{Helton2007, Mendels2007}.
A spinon-like continuum has been reported in herbertsmithite at high energies ($> 2$ meV) from neutron experiments \cite{Han2012}, and a small spin gap ($\sim 1$ meV) has been reported at low energies by NMR \cite{Fu2015}.
Nevertheless, it has been pointed out that excess Cu$^{2+}$ ions replace nonmagnetic Zn$^{2+}$ ions between the kagom\'e layers by ~15\% \cite{Freedman2010, Fu2015}, obscuring whether an ideal two-dimensional KHA is realized in herbertsmithite.

Searching for an ideal KHA has led to recent discoveries of new kagom\'e materials from structural polymorphs of herbertsmithite, such as Zn kapellasite ZnCu$_3$(OH)$_6$Cl$_2$ \cite{Krause2006}, haydeeite $\alpha$-MgCu$_3$(OH)$_6$Cl$_2$ \cite{Schluter2007}, Cd kapellasite CdCu$_3$(OH)$_6$(NO$_3$)$_2\cdot$H$_2$O \cite{Okuma2017}, and Ca kapellasite CaCu$_3$(OH)$_6$Cl$_2\cdot$0.6H$_2$O \cite{HYoshida2017}.
Whereas the Zn ions in herbertsmithite are located between the kagom\'e layers, cations in these kapellasites or haydeeite are in the same kagom\'e layer of Cu ions, resulting in smaller coupling between the kagom\'e layers and hence better two-dimensionality \cite{Janson2008}.
Among these new kagom\'e materials, Ca kapellasite (Fig.~1(a) and (b)) is the most promising KHA candidate.
In Zn kapellasite and haydeeite, the nearest-neighbor interaction is found to be ferromagnetic \cite{Bernu2013, Jeschke2013}.
In contrast, the fitting of the temperature dependence of the magnetic susceptibility \cite{HYoshida2017} shows that the spin Hamiltonian in Ca kapellasite is well approximated as an ideal KHA (Fig.~1(a), $J_1=52.2$~K, $J_2=-6.9$~K, $J_d=11.9$~K), which is supported by a first-principles calculation \cite{JeschkePrv} showing a very dominant $J_1$ (($J_1$, $J_2$, $J_d$) = (64, 2.8, -2.0) for onsite repulsion $U=7.0$~eV).
Further, because the ionic radii of Ca$^{2+}$ ion (1.0~\AA) is much larger than that of Cu$^{2+}$ (0.73~\AA), there are no site mixings in Ca kapellasite, in contrast to Zn/Cu (Mg/Cu) site mixings of $\sim 27$\% ($\sim 16$\%) in Zn kapellasite
\cite{Kermarrec2014} (haydeeite \cite{Colman2010}).

The temperature dependence of the magnetic susceptibility of Ca kapellasite features a broad peak at $\sim 30$~K, indicating the development of a short-range spin correlation, followed by a peak at $T^*=7.4$~K \cite{HYoshida2017}. 
The temperature dependence of the heat capacity also features a peak at $T^*$, indicating a magnetic transition at $T^*$.
This magnetic transition was confirmed in NMR, reporting a small broadening of the spectrum of $\sim 0.05 \mu_B$ below $T^*$ \cite{Ihara2017}.
Although the magnetic ground state of Ca kapellasite is not a disordered state, the ordering temperature is much smaller than $J_1$, suggesting that a spin liquid state is realized over a wide temperature range, $T^* < T < J_1/k_B$.
A zero-temperature extrapolation of the heat capacity data above $T^*$ reveals a large residual in the linear $T$ term, $\gamma$, implying the presence of gapless spin excitations in the spin liquid phase.
Also a finite $\gamma$ is suggested to remain even in the ordered phase below $T^*$ \cite{HYoshida2017, Ihara2017}, implying the presence of unusual gapless spin excitations.

To study the elementary excitation in Ca kapellasite, we performed longitudinal ($\kappa_{xx}$) and transverse ($\kappa_{xy}$) thermal conductivity measurements (see Fig.~1(c) and the Supplemental Material (SM) \cite{SM} for details).
Itinerant excitations are studied by measuring $\kappa_{xx}$ down to very low temperatures and the spin chirality of the underlying spin system can be probed by measuring $\kappa_{xy}$.
Thermal Hall measurements have been used in ferromagnetic insulators \cite{Onose2010, Ideue2012, HirschbergerPRL2015} and are well explained by the Berry phase effect of the magnon bands \cite{Katsura2010,Matsumoto2014}.
Thermal Hall effects have also been reported in spin liquid states such as the quantum spin ice material Tb$_2$Ti$_2$O$_7$ \cite{HirschbergerScience2015}, the kagom\'e antiferromagnet volborthite \cite{Watanabe2016}, and the Kitaev candidate $\alpha$-RuCl$_3$ \cite{Kasahara2017}.
In volborthite \cite{Watanabe2016}, $\kappa_{xy}$ emerges as entering the spin liquid region and shows a sign change near the N\'eel temperature, suggesting that the thermal Hall effect has a spin origin. Also, the development of spin correlations in the spin liquid phase is important for $\kappa_{xy}$.
However, the detailed origin of $\kappa_{xy}$ has yet to be understood.
In this Letter, we report a distinct thermal Hall signal in Ca kapellasite and reveal that both the temperature dependence and the magnitude of $\kappa_{xy}$ is understood by the thermal Hall effect of Schwinger bosons described by the Schwinger-boson mean-field theory (SBMFT) \cite{Hyunyong2015}.


\begin{figure}[!tbh]
\centering
\includegraphics[width=\linewidth]{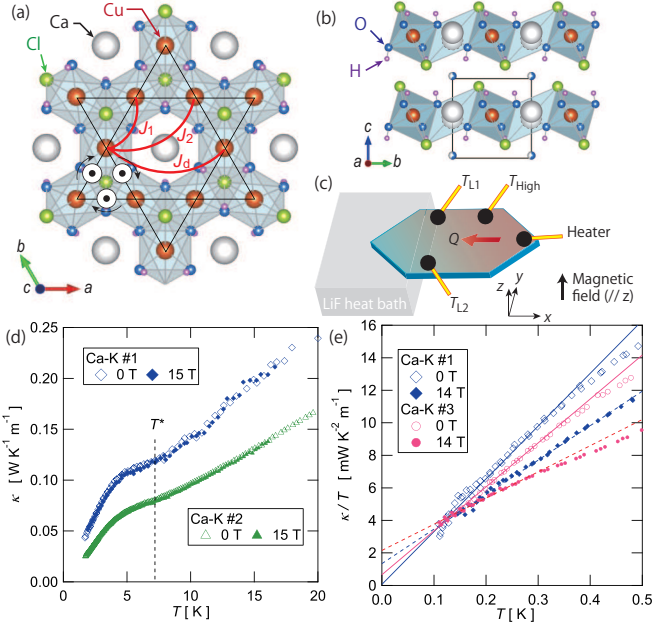}
\caption{
(a, b) Crystal structure of Ca kapellasite viewed along the $c$ axis (a) and the $a$ axis (b).
The direction of the Dzyaloshinskii--Moriya interaction ($D_{ij}$) considered in Eq. (1) is shown by $\odot$ symbols.
The black arrows next to the $\odot$ symbols indicate the direction in the difinition of $D_{ij}$ (See the SM \cite{SM}).
$J_1$, $J_2$ and $J_d$ (the solid red lines) represent the nearest-neighbor, next nearest-neighbor, and diagonal magnetic interactions, respectively. Ca$^{2+}$ ions are randomly located in either of two Wyckoff positions that are slightly different along the $c$ axis (the solid and dotted circles in (b) \cite{HYoshida2017}).
(c) An illustration of our experimental setup. Three thermometers ($T_{\rm High}$, $T_{\rm L1}$, $T_{\rm L2}$) and a heater were attached to the sample. A heat current $Q\parallel x$ was applied within the $ab$ plane and a magnetic field was applied along the $c\parallel z$ axis. See the SM \cite{SM} for details.
(d, e) Temperature dependence of $\kappa_{xx}$ for samples \#1 (the blue diamonds) and \#2 (the green triangles) at high temperatures at 0 and 15~T (d) and that of $\kappa_{xx}/T$ for samples \#1 and \#3 (the pink circles) below 0.5 K at 0 and 14~T (e). The slope of $\kappa_{xx}(T)$ is decreased below $T^*=7.4$~K. 
The zero-temperature extrapolation of $\kappa_{xx}/T$ (the solid and dashed lines for 
0 and 14 T, respectively) shows zero or negligibly small residual.
}
\label{fig1}
\end{figure}


The temperature dependence of $\kappa_{xx}$ above 2~K and that of $\kappa_{xx}/T$ below 0.5~K are shown in Fig.~1(d) and (e), respectively.
The magnitude of $\kappa_{xx}$ for Ca kapellasite is found to be very low, at about an order of magnitude below that for volborthite \cite{Watanabe2016}.
The longitudinal thermal conductivity of an insulator is given by $\kappa_{xx}= \kappa_{xx}^{ph} + \kappa_{xx}^{sp}$, where $\kappa_{xx}^{ph}$ ($\kappa_{xx}^{sp}$) is $\kappa_{xx}$ by phonons (spins).
As seen in Fig.~1(d) and in the SM \cite{SM}, the magnetic field strength has little effect on $\kappa_{xx}$ above $T^*$.
This is in contrast to the decrease with magnetic field observed in volborthite attributed to a field suppression effect on $\kappa_{xx}^{sp}$ \cite{Watanabe2016}.
Therefore, above $T^*$, phonons provide the dominant contribution to $\kappa_{xx}$.
Below $T^*$, the slope of $\kappa_{xx}(T)$ slightly decreases, followed by a shoulder at $\sim 5$~K. This suggests that the magnetic order at $T^*$ increases $\kappa_{xx}^{ph}$ and/or an additional magnon thermal conduction appears below $T^*$.

Below ~0.3 K, $\kappa_{xx}$ shows a $T^2$ temperature dependence (Fig.~1(e)). 
A small quadratic $\kappa_{xx}$ is often observed in amorphous glass materials \cite{Pohl2002} and in single crystals with randomness because of site mixings \cite{Xu2016, Sugii2017}.
Therefore, we conclude that a similar phonon-glass state is realized in Ca kapellasite because of the random distribution of Ca$^{2+}$ ions and/or the non-stoichiometric composition of H$_2$O molecules.
Although the field dependence of $\kappa_{xx}$ at low temperatures shows sizable $\kappa_{xx}^{sp}$ in $\kappa_{xx}$ \cite{SM}, the linear extrapolation of $\kappa_{xx}/T$ with respect to $T$ shows zero or negligibly small residual up to $1 \sim 2$~mW~K$^{-2}$~m$^{-1}$ (the solid and dashed lines in Fig.~1(e))
compared to the values found in other quantum spin-liquid candidates~\cite{MYama2010, Shimozawa2017}, indicating the absence of itinerant gapless excitations with linear-$T$ dependence.
This suggests that the gapless excitations obtained from heat capacity \cite{HYoshida2017} and NMR \cite{Ihara2017} measurements are localized and do not contribute to $\kappa_{xx}$.

\begin{figure}[!tbh]
\centering
\includegraphics[width=\linewidth]{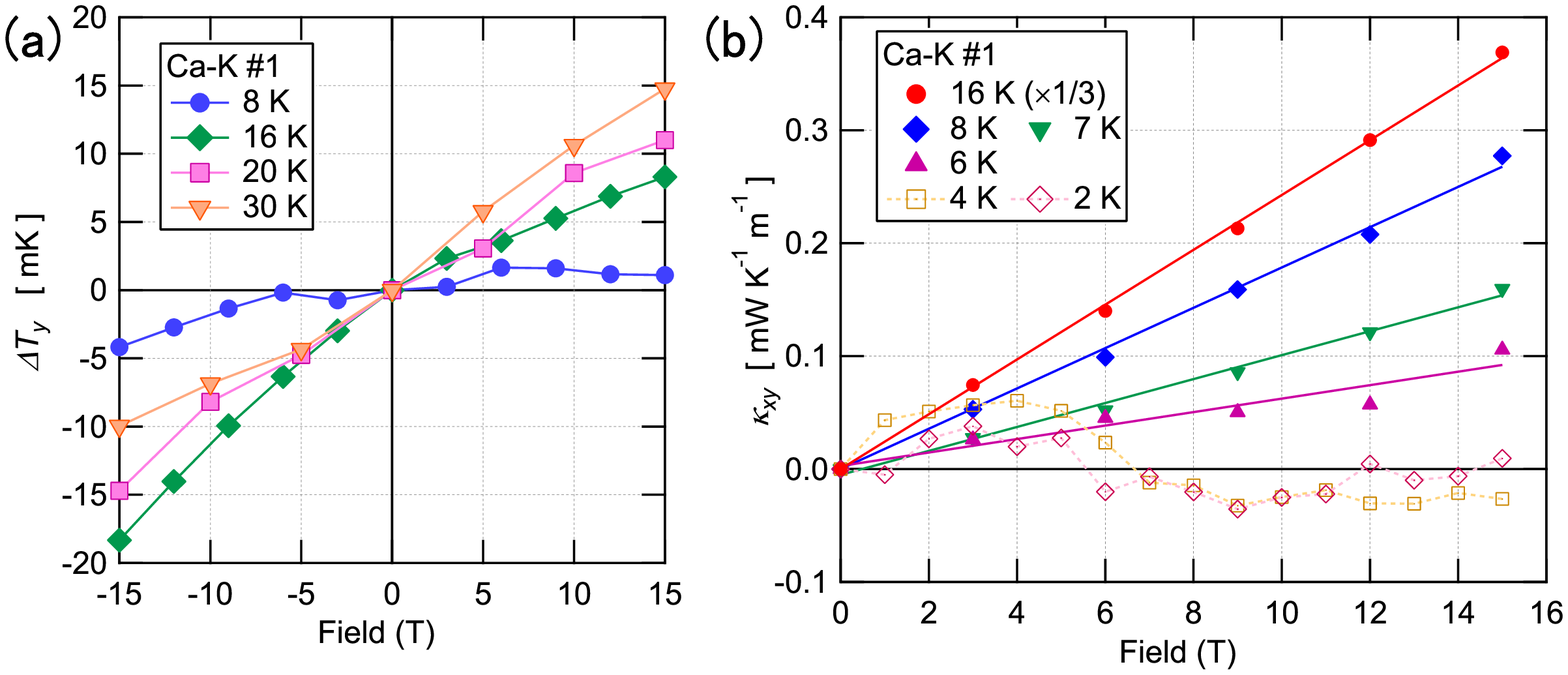}
\caption{
The field dependence (a) of the transverse temperature difference  $\Delta T_y (B)$ and (b) of $\kappa_{xy}(B)$. Solid lines in (b) represent linear fits.
The field dependence of $\kappa_{xy}(B)$ at other temperatures is shown in the SM~\cite{SM}.
}
\label{dTy_kxy}
\end{figure}

The field dependence of the transverse temperature difference ($\Delta T_y (B) \equiv T_{\rm L1}(B) - T_{\rm L2}(B)$)
shows a large asymmetric field dependence (Fig.~\ref{dTy_kxy}(a)).
The symmetric field dependence of $\Delta T_y (B)$ with respect to the field direction is caused by the field dependence of $\kappa_{xx}$ included in $\Delta T_y (B)$ by a misalignment of the thermal contacts.
We confirmed that the asymmetric component of $\Delta T_y (B)$ increases linearly with respect to the heater power $Q$ (Fig.~S3 in the SM), indicating the presence of the thermal Hall signal in Ca kapellasite.
By antisymmetrizing $\Delta T_y (B)$, we estimated the field dependence of $\kappa_{xy}$.
The field dependence of $\kappa_{xy}(B)$ (Fig.~\ref{dTy_kxy}(b)) is linear above $T^*$.
Below $T^*$, the field dependence of $\kappa_{xy}(B)$ is non-linear; $\kappa_{xy}(B)$ shows a peak around 4~T and becomes negligibly small above 6~T (Fig.~\ref{dTy_kxy}(b)). This different field dependence below $T^*$ suggests that the magnetic order has a significant effect on the elementary excitations producing the thermal Hall signal. A change in magnetic structure below $T^*$ has also been inferred at 6~T from the change in slope of the magnetization \cite{HYoshida2017} that may be related to the disappearance of $\kappa_{xy}(B)$ above 6~T.


\begin{figure}[!tbh]
\centering
\includegraphics[width=0.9\linewidth]{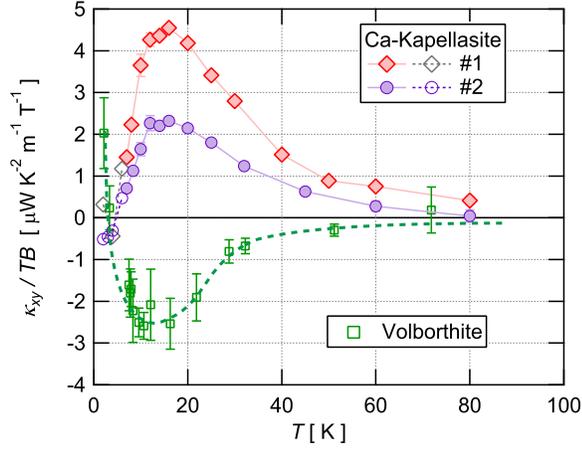}
\caption{
The temperature dependence of $\kappa_{xy}/TB$ of Ca kapellasite (samples \#1 and \#2) and that of volborthite \cite{Watanabe2016}. The filled (open) symbols represent data estimated by the linear fit of $\kappa_{xy}$ (data at 15~T). The data of volborthite is taken from Ref.~\cite{Watanabe2016}. The error bars correspond to one standard deviation and are of the same order as the size of the symbol or smaller for data of Ca kapellasite. 
}
\label{fig_k_xy_T}
\end{figure}

From the linear fit for $\kappa_{xy}(B)$ (the straight lines in Fig.~\ref{dTy_kxy}(b)), we estimated the slope $\kappa_{xy}/B$ at each temperature and plotted the temperature dependence of $\kappa_{xy}/TB$ (filled symbols in Fig.~\ref{fig_k_xy_T}).
Below $T^*$, we estimated $\kappa_{xy}/TB$ from $\kappa_{xy}$ at 15~T (open symbols in Fig.~\ref{fig_k_xy_T}).
We note that $\kappa_{xy}/TB$ data below $T^*$ is shown for reference owing to the non-linear field dependence of $\kappa_{xy}$.
Clarifying $\kappa_{xy}$ below $T^*$, which requires the detail of the magnetic order, remains as a future work as discussed later.
We find that $\kappa_{xy}/TB$ for both Ca kapellasite samples exhibit virtually the same temperature dependence.
The magnitudes of $\kappa_{xy}$ for both differ by a factor of $\sim 2$,
which is mostly attributed to the ambiguity in the estimation of the sample geometry (see the SM~\cite{SM} for more details).
As seen in Fig.~\ref{fig_k_xy_T}, $\kappa_{xy}/TB$ increases as the temperature is lowered, then peaks at $\sim 20$~K followed by a rapid decrease to zero below $T^*$.
This temperature dependence, in particular the peak in $|\kappa_{xy}/TB|$, is almost the same with that of volborthite.
Remarkably, the absolute value of $\kappa_{xy}/TB$ of Ca kapellasite is also similar to that of volborthite, whereas $\kappa_{xx}$ of Ca kapellasite is about one order of magnitude smaller than that of volborthite.
Because $\kappa_{xx}$ is dominated by phonons in this temperature range, similar $|\kappa_{xy}/TB|$ magnitudes in these kagom\'e compounds with different $\kappa_{xx}$ magnitudes suggests that the thermal Hall effect does not come from phonons \cite{Sugii2017}.
Given almost the same magnitude for the effective spin interaction energy $J/k_B \sim 60$~K of the two compounds, similar $\kappa_{xy}/TB$ implies the presence of a common thermal Hall effect from spin excitations of the kagom\'e antiferromagnets.


To investigate the origin of $\kappa_{xy}$, we have simulated $\kappa_{xy}$ adopting the SBMFT \cite{ArovasAuerbach1988} for KHA with the Dzyaloshinskii--Moriya (DM) interaction, which reads
\begin{equation}
\mathcal{H} = \frac{1}{2} \sum_{\langle i,j \rangle} 
 \left(
	J\,\v S_i \cdot \v S_j + 
	D_{ij}\, \v S_i \times \v S_j \cdot \hat{z}
 \right)
 - g \mu_B \sum_i \v B \cdot \v S_i,
\label{eq:H}\end{equation}
where $D_{ij}$ is the DM interaction, $g$ the $g$ factor, $\mu_B$ the Bohr magneton, and the direction of the magnetic field $\v B$ aligns with the $z$ axis. 
The sign of $D_{ij}$ is assumed to be positive if $i \rightarrow j$ is clockwise direction from the center of each triangle plaquette in the kagom\'e lattice, and we define $D_{ij} = - D_{ji} = D$.
SBMFT has been employed to study the possible spin liquid ground states and the excitations of quantum antiferromagnets \cite{ArovasAuerbach1988, Messio2017,Sachdev1992,WangVishwanath2006,LiErnChern2017,Yang16,Wang10,Sungbin14,Messio13}.
In the SBMFT framework, spin is expressed by a pair of bosons $(b_{i\uparrow}, b_{i\downarrow})$ as 
$\v S_i = \frac{1}{2}\sum_{\alpha,\beta=\uparrow,\downarrow} b^\dagger_{i\alpha} \bm{\sigma}_{\alpha \beta} b_{i\beta}$,
where $\bm{\sigma}$ is the Pauli matrices.
We decouple the Hamiltonian by taking a mean-field value of the bond operator $\chi_{ij} = \langle b^\dagger_{i\sigma} b_{j \sigma} \rangle$ and diagonalize it to find the energy bands.
Because of the nature of the DM interaction, $\chi_{ij}$ is a complex number, and therefore the energy bands are gapped. Each band now carries a different Berry flux, and this is directly related to the thermal Hall conductivity through the relation \cite{Katsura2010, Matsumoto2014}:
\begin{equation}
\kappa_{xy}^{\rm SBMF} = 
 - \frac{k_B^2 T}{\hbar N_t} \sum_{\bm{k},n,\sigma}
 \left[
    c_2\left( n_B\left(\frac{E_{n\v k \sigma}}{k_B T}\right) \right) - \frac{\pi^2}{3}
 \right]
\Omega_{\bm{k}n \sigma},
\label{eq:kxy_SB}\end{equation}
where $c_2$ is a distribution function of the Schwinger bosons, $n_B$ the Bose-Einstein distribution function, $E_{\bm{k}n \sigma}$ the energy eigenvalue, and $\Omega_{\bm{k}n \sigma}$ the Berry curvature (see the SM \cite{SM} for details).
Equation~(\ref{eq:kxy_SB}) can be expressed as $\kappa_{xy}^{\rm SBMF}/T = (k_B^2/\hbar)f_{\rm SBMF}(k_B T/J, D/J, g \mu_B B/J)$, where $f_{\rm SBMF}$ is a dimensionless function.
Given that $\kappa_{xy}$ is an odd function of both $D$ and $B$, one has the approximation $\kappa_{xy}^{\rm SBMF}/T = (k_B^2/\hbar)(D/J)(g \mu_B B/J)\tilde{f}_{\rm SBMF}(k_B T/J)$ when both $D$ and $g \mu_B B$ are smaller than $J$.

\begin{figure}[!tbh]
\centering
\includegraphics[width=0.9\linewidth]{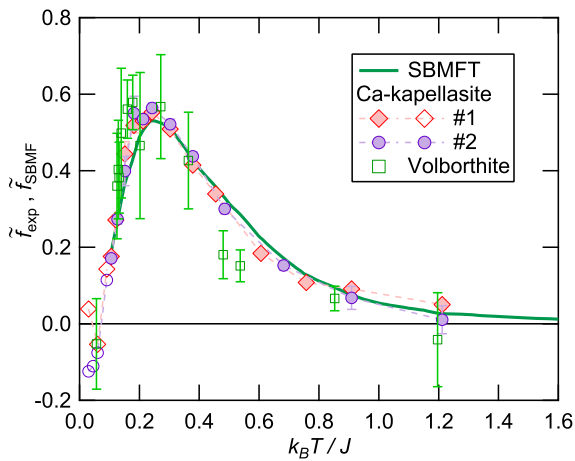}
\caption{
Dimensionless thermal Hall conductivity $\tilde{f}_{\rm exp}(k_B T/J)$ and $\tilde{f}_{\rm SBMF}(k_B T/J)$. The solid line represents numerical results obtained using the Schwinger-boson mean-field theory of $D/J=0.1$ (see the SM \cite{SM} for more results). The values of parameters ($J/k_B$, $D/J$) used for the experimental data are (66, 0.12), (66, 0.06), and (60, -0.07) for Ca kapellasite \#1, \#2, and volborthite, respectively.
}
\label{fig_k_xy_2D_T}
\end{figure}

Our central finding is that all $\kappa_{xy}(T)$ data (Fig.~\ref{fig_k_xy_T}) are well fitted by $\tilde{f}_{\rm SBMF}(k_B T/J)$ calculated by our SBMFT for the kagom\'e Hamiltonian, Eq.~\ref{eq:H}.
To compare the numerical results with experiment, we estimate the dimensionless $\tilde{f}_{\rm exp}(k_B T/J)$ using $\tilde{f}_{\rm exp} \equiv \kappa_{xy}^{2D} / (k_B^2 TD g \mu_B B / \hbar J^2)$ as a function of $k_B T/J$, where $\kappa_{xy}^{2D} = \kappa_{xy} d$ is $\kappa_{xy}$ of one kagom\'e layer, $d=5.76$ (7.22) {\AA} the interlayer distance of Ca kapellasite \cite{HYoshida2017}(volborthite \cite{HYoshida2012}), and $g = 2.14$ (2.28) the $g$ factor of Ca kapellasite (volborthite \cite{Ishikawa2015}).
Remarkably, by choosing only $J$ and $D$ as fitting parameters, we find that $\tilde{f}_{\rm exp}(k_B T/J)$ for all kagom\'e antiferromagnets and that of the SBMFT simulations converge to one single curve (Fig.~\ref{fig_k_xy_2D_T}).
This excellent agreement, both qualitative and quantitative, demonstrates that the thermal Hall effect in these kagom\'e antiferromagnets derives from spins in the spin liquid phase.
Figure~\ref{fig_k_xy_2D_T} further verifies not only that both the kagom\'e materials are well fitted by the simple Hamiltonian of Eq.~(1), but that $\kappa_{xy}$ in a kagom\'e spin liquid is well described by a simple one-variable scaling function $\tilde{f}_{\rm SBMF}(k_B T/J)$, thereby capturing the nature of $\kappa_{xy}(T)$ for KHA.
The values of $J/k_B=66$~K (60 K) for Ca kapellasite (volborthite) are in good agreement with the values estimated by the temperature dependence of the magnetic susceptibility.
The values of $|D/J|$ for both materials are also similar to the value estimated from the deviation of the $g$ factor from 2 (see Table S1 in the SM~\cite{SM} summarizing $J$ and $D$ values obtained by different methods).
Therefore, the fitting results of $J$ and $D$ support arguments for the common temperature dependence of $\kappa_{xy}$.
We note that the factor 2 difference of the $D$ values for Ca kapellasite simply reflects the ambiguity in estimating the absolute value of $\kappa_{xy}$ which is mainly caused by the irregular shape of the crystal~\cite{SM}.


The good agreement between the observed $\kappa_{xy}$ and the SBMFT simulation shows that the spin liquid state of these kagom\'e antiferromagnets is well described by the $U(1)$ spin liquid with bosonic spinons. 
Whether our ansatz is the only successful state for describing $\kappa_{xy}(T)$ or other spin liquid states, in particular spin liquids with fermionic spinons having a different $\kappa_{xy}(T)$, remains an open question.
The thermal Hall effect calculated for the $120^\circ$ ordered state in a kagom\'e antiferromagnet has been shown to have a different temperature dependence \cite{OwerreEPL2017}.
Therefore, the short-ranged magnons of a kagom\'e antiferromagnet that may survive above $T^*$ can be excluded as excitations producing the observed $\kappa_{xy}$.

The spin Hamiltonian of volborthite at low temperatures has been shown to deviate from the ideal KHA \cite{Ishikawa2015} but has been suggested as an effective triangular model of coupled trimers with an effective interaction energy of $\sim 35$~K obtained from a first-principles calculation \cite{Janson2016}.
Because the deviation of $\kappa_{xy}$ of volborthite from our SBMFT simulation based on the KHA was not clearly observed, the effects of a modified triangular trimer model might be confined to lower temperatures.

The fitting parameters ($J$ and $D$) obtained by scaling $\kappa_{xy}$ to $\tilde{f}_{\rm SBMF}$ are the two most decisive parameters in finding the ground state of a KHA.
This ground state has been studied using the Schwinger boson approximation \cite{Sachdev1992, WangVishwanath2006, LiErnChern2017, Messio2017} and shows that magnetically ordered states appear through the Bose--Einstein condensation of the Schwinger bosons.
Therefore, the ordered state in Ca kapellasite below $T^*$ may be described by a similar condensation of the Schwinger bosons.
According to a recent numerical study using SBMFT with a DM interaction \cite{Messio2017}, a chiral spin liquid phase is suggested to be stabilized for a $D/J$ value that is close to our results concerning Ca kapellasite.
Although the detailed magnetic structure of the ordered state below $T^*$ has yet to be clarified, the ground state of Ca kapellasite may be located near the chiral spin liquid phase.
We note that our present SBMFT simulation is not applicable at very low temperatures near the N\'eel order because we omit $b_{i \sigma} b_{j \sigma}$ terms in our calculation.
For the Schwinger bosons, these terms describe pairing interactions and play an important role near the ordering temperature in determining the ground state.
An exact description of the ordered state by including these terms to our mean-field approximation remains as a future work.

In summary, we have observed a distinct thermal Hall effect in a kagom\'e antiferromagnet Ca kapellasite. The thermal Hall conductivity of both Ca kapellasite samples exhibit a similar temperature dependence as $\kappa_{xy}$ of another kagom\'e antiferromagnet volborthite. In a simulation using SBMFT, we find all observed $\kappa_{xy}$ values converge onto a single curve by adjusting only $J$ and $D$, suggesting that the thermal Hall conductivity of a kagom\'e antiferromagnet has a common temperature dependence described by Schwinger bosons.

\begin{acknowledgments}
We thank Shunsuke Furukawa, Yoshihiko Ihara, Kazuki Iida, Harald O. Jeschke, Yong Baek Kim, Tsutomu Momoi, Masaki Oshikawa, and Oleg Starykh for fuitful discussions. 
This work was supported by Yamada Science Foundation, Toray Science Foundation, and KAKENHI (Grants-in-Aid for Scientific Research) Grant No. 15K17686, No. 15K17691, No. 17K18747, and No. 18K03529.
H.L. and N.K. were supported by MEXT as "Exploratory Challenge on Post-K computer" (Frontiers of Basic Science: Challenging the Limits).
J. H. H. was supported by Samsung Science and Technology Foundation under Project No. SSTF-BA1701-07.

H.D., M.A., and H.L. contributed equally to this work.

\end{acknowledgments}

%


\clearpage
\onecolumngrid

\appendix
\vspace{15pt}

\fontsize{12pt}{1cm}\selectfont

\begin{center}
{\Large \bf ---Supplemental Material For ``Spin Thermal Hall Conductivity of a Kagom\'e Antiferromagnet"---}
\end{center}

\setcounter{figure}{0}
\setcounter{equation}{0}
\setcounter{table}{0}
\renewcommand{\thefigure}{S\arabic{figure}}
\renewcommand{\theequation}{S\arabic{equation}}
\renewcommand{\thetable}{S\Roman{table}}
\baselineskip=6mm

\noindent{\bf S1.\ \ \  Materials and Methods}
\vspace{1em}

The single crystals of Ca kapellasite were synthesized by a hydrothermal method as described in ref.~\cite{HYoshida2017}.
Typical sample size is $\sim 1$~mm $\times$ $\sim 1$~mm $\times$ $\sim 0.1$~mm.

\begin{figure}[htbp]
\centering
\includegraphics[width=0.8\linewidth]{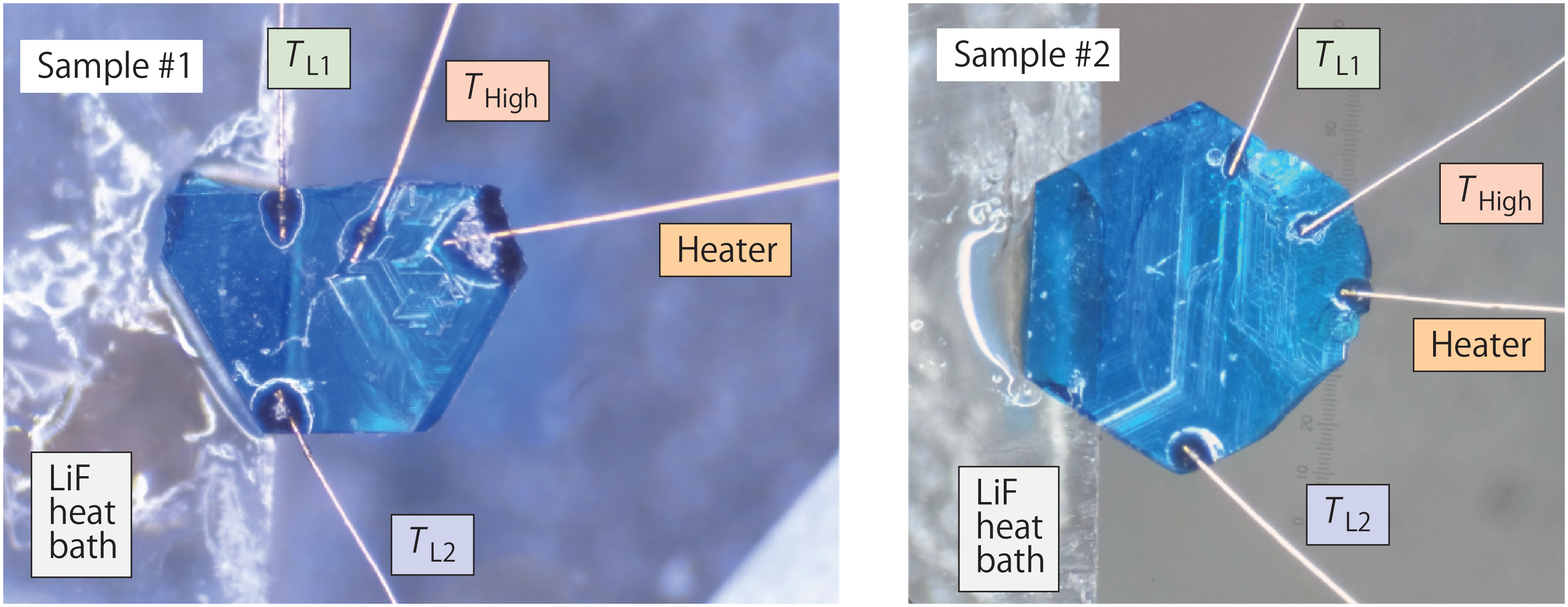}
\caption{
Pictures of sample \#1 (left) and sample \#2 (right) setup. Gold wires (25 $\mu$m diameter) connected to the three thermometers ($T_{\rm High}$, $T_{\rm L1}$, $T_{\rm L2}$) and a heater were fixed to the sample by stycast 2850FT epoxy.}
\label{pic}
\end{figure}

Pictures of the experimental setup (sample \#1 (left) and sample \#2 (right)) are shown in Fig.~\ref{pic}.
A heat current $Q\parallel x$ was applied within the $ab$ plane and a magnetic field was applied along the $c\parallel z$ axis.
Three thermometers ($T_{\rm High}$, $T_{\rm L1}$, $T_{\rm L2}$) were attached so that both the longitudinal ($\Delta T_x (B) \equiv T_{\rm High} (B) - T_{\rm L1} (B)$) and the transverse ($\Delta T_y (B) \equiv T_{\rm L1} (B) - T_{\rm L2} (B)$) thermal gradients can be simultaneously measured.
Cernox (RuO$_2$) thermometers were used for measurements at high temperatures above 2 K (low temperatures below 2 K).
All thermometers were carefully calibrated in magnetic fields by calibrated thermometers placed outside the magnet.
For Cernox thermometers used for the thermal Hall measurements, we confirmed that all three thermometers show the identical magnetoresistance with the value reported in the previous report~\cite{Brandt1999} within our resolution.
To avoid a background Hall signal from metals, an insulating single crystal of LiF was used as a heat bath to which the sample was attached by a small amount of Apiezon grease.

The longitudinal thermal conductivity ($\kappa_{xx}$) and the thermal Hall conductivity ($\kappa_{xy}$) are obtained by
\begin{equation}
\frac{1}{wt} 
\begin{pmatrix}
Q \\
0
\end{pmatrix}
=
\begin{pmatrix}
\kappa_{xx}(B) & \kappa_{xy}(B) \\
-\kappa_{xy}(B) & \kappa_{xx}(B)
\end{pmatrix}
\begin{pmatrix}
\Delta T_x (B)/L \\
\Delta T_y^{Asym} (B) / w
\end{pmatrix},
\end{equation}
where $\Delta T_y^{Asym} (B) \equiv \left[\Delta T_y (+B) - \Delta T_y (-B)\right] / 2$, $t$ is the thickness of the sample, and $L$ ($w$) is the distance between the thermal contacts for $T_{\rm High}$ and $T_{\rm L1}$ ($T_{\rm L1}$ and $T_{\rm L2}$).

Here, we note about an error in estimating the absolute value of $\kappa_{xx}$ and that of $\kappa_{xy}$ owing to the irregular shape of the crystals.
Whereas we can accurately control the heat current $Q$ and can measure the temperature differences $\Delta T_x$ and $\Delta T_y$ in high precision,  there is relative large ambiguity in estimating the sample geometries; the shapes of the Ca kapellasite crystals are irregular with non-uniform thickness and width, and the sample size is not long enough to ignore the size effect of the thermal contacts to the gold wires (see Fig.~\ref{pic}). As a result, the ambiguity of $\sim 20$\% for each dimension is unavoidable.
This ambiguity effect is larger in the estimation of $\kappa_{xy}$ which, in a worst case, can sum up to the ambiguity of factor of 2 ($\sim 1.15^5$) in the estimation of $\kappa_{xy}$.

The difference of $\kappa_{xy}$ between sample \#1 and sample \#2 is thus mainly caused by this ambiguity, although we cannot exclude the possibility of the sample dependence.
The different ratio between $D/J$ of sample \#1 and that of sample \#2 obtained by the fitting (Fig.~4 in the main text) is also caused by this error.

\vspace{1em}
\vspace{1em}
\noindent{\bf S2.\ \ \  Field dependence of $\kappa_{xx}$}
\vspace{1em}

\begin{figure}[htbp]
\centering
\includegraphics[width=\linewidth]{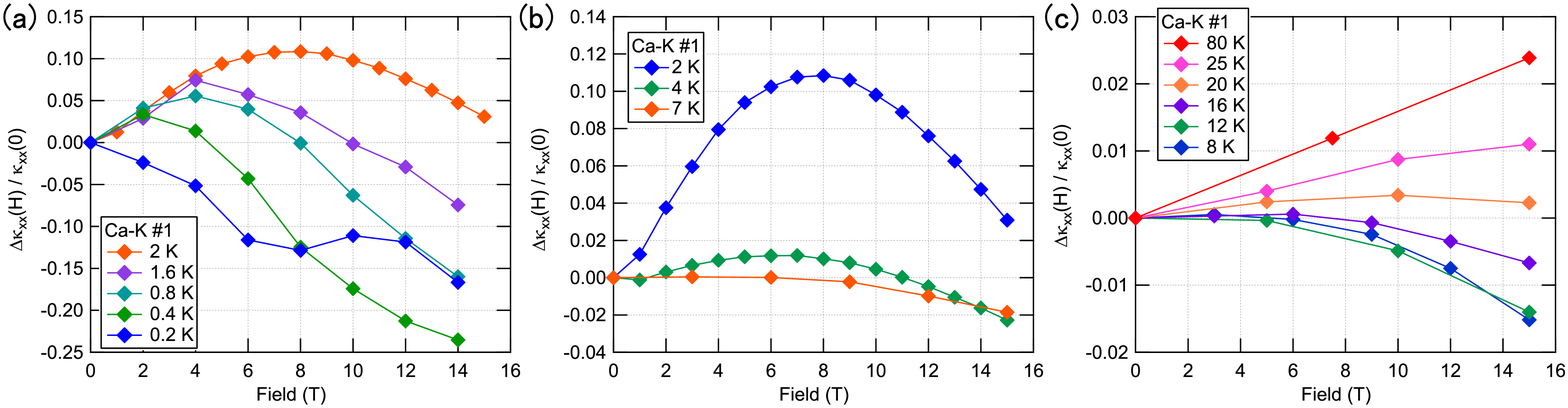}
\caption{
The field dependence of $\kappa_{xx}$ below 2 K (a), for 2--7 K (b), and above 8 K (c). The vertical axis shows a difference by field, $\Delta \kappa_{xx} (B) \equiv \kappa_{xx} (B) - \kappa_{xx} (0)$, normalized by the zero field value $\kappa_{xx} (0)$ at each temperature.
}
\label{kxx_H}
\end{figure}

The field dependence of $\kappa_{xx}$ at different temperatures is shown in Figs. \ref{kxx_H}(a)-(c).
At the lowest temperature (Fig. \ref{kxx_H}(a)), $\kappa_{xx}$ is decreased by magnetic field.
This is opposite to the field dependence of the phonon thermal conduction ($\kappa_{xx}^{ph}$) which is usually expected to be increased by magnetic field because the magnetic scatterings for phonons are suppressed by magnetic field.
A resonant phonon scattering with magnetic impurities, which is known to decrease $\kappa_{xx}^{ph}$ under magnetic field \cite{Berman, Watanabe2016}, is not the origin of the field suppression effect. The field dependence of the resonant scattering is a function of $x^4 e^x/(e^x-1)^2$, where $x=g\mu_B B/k_B T$ and $g$ is the $g$-factor. The effect of the resonant scattering is thus only effective at $g\mu_B B \sim 4k_B T$ which is $\sim 0.6$ T at 0.2 K for $g_c=2.14$ \cite{HYoshida2017}, whereas the field suppression effect persists up to 14 T.
Therefore, the decrease of $\kappa_{xx}$ by magnetic field at 0.2 K can be attributed to a field suppression effect on the spin thermal conduction ($\kappa_{xx}^{sp}$), showing that at least 20\%
of $\kappa_{xx}$ is provided by $\kappa_{xx}^{sp}$ at low temperatures.

At elevated temperatures, $\kappa_{xx}$ increases at lower fields (Fig.~\ref{kxx_H} (a) and (b)).
This increase becomes larger at higher temperatures, which can be attributed to the field enhancing effect on $\kappa_{xx}^{ph}$.
At higher fields, the field suppression effect becomes dominant, resulting in a peak at 8 T at 2 K (Figs.~\ref{kxx_H}(a) and (b)).
The peak structure becomes smaller above 2 K and the field suppression effect becomes dominant again at $T\sim T^*$ (Fig.~\ref{kxx_H} (b)). 
Above 20 K (Fig.~\ref{kxx_H}(c)), only the field enhancing effect is observed, which is consistent with the fact that $\kappa_{xx}^{ph}$ becomes larger at higher temperatures.

We note that the field dependence of $\kappa_{xx}$ below $T^*$ is consistent either the two scenarios that the decrease of the slope of $\kappa_{xx}$ at $T^*$ is caused by either an increase of the $\kappa_{xx}^{ph}$ or an additional magnon thermal conduction.

Phonons are scattered by both intrinsic and impurity spins by spin-phonon scatterings. The magnetic order at $T^*$ aligns the intrinsic spins, giving rise to the increase of $\kappa_{xx}^{ph}$ by the decrease of the spin-phonon scatterings. This magnetic order also enhances the field effect on $\kappa_{xx}^{ph}$ because the spin-phonon scatterings below $T^*$ are dominated by the impurity spins which can be easily aligned by magnetic field.
Therefore, the former scenario is consistent with the larger field enhance effect observed below $T^*$ (Fig. S2(a) and (b)).

The latter is consistent with the $\kappa_{xx}^{sp}$ observed by the field suppression effect on $\kappa_{xx}$ at the lowest temperature (Fig. S2(a)).

\vspace{1em}
\vspace{1em}
\noindent{\bf S3.\ \ \  Additional data of the thermal Hall measurements}
\vspace{1em}

\begin{figure}[htbp]
\centering
\includegraphics[width=0.5 \linewidth]{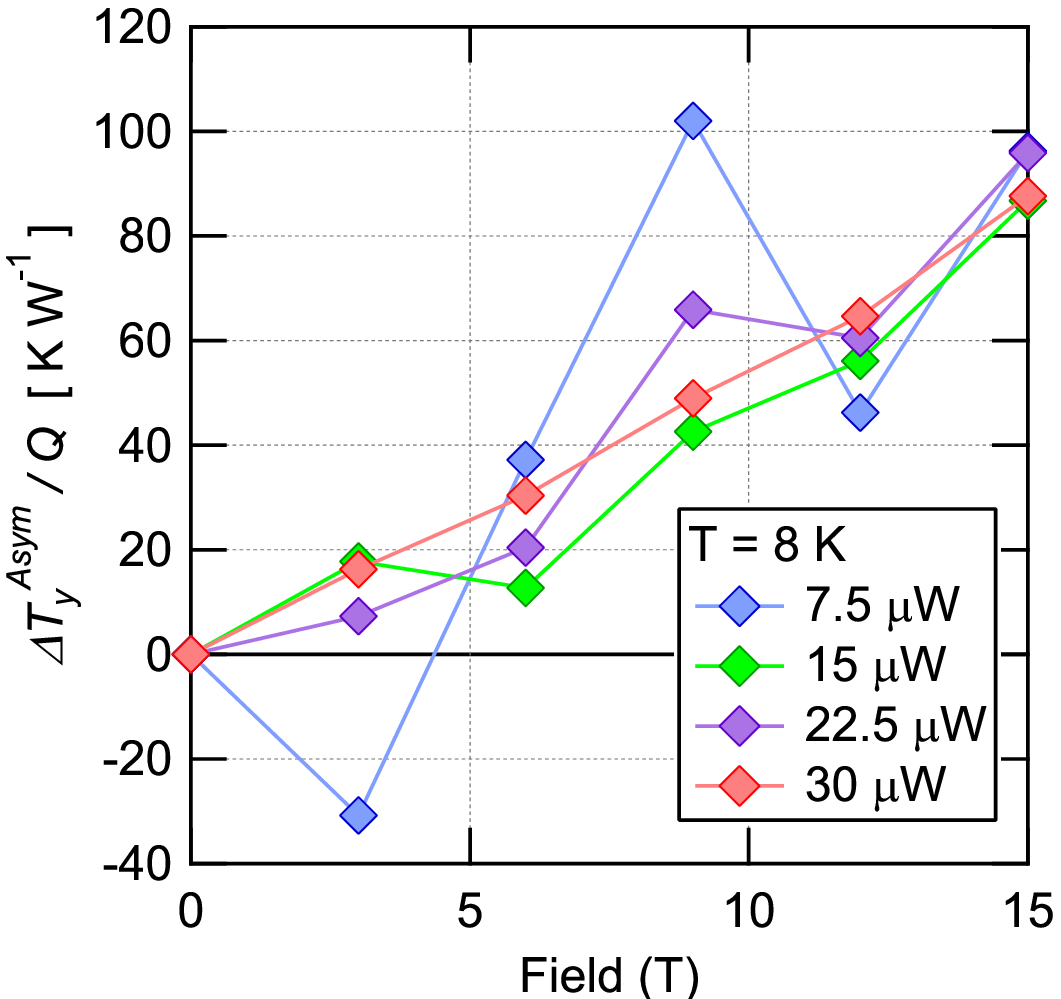}
\caption{
The field dependence of $\Delta T_y^{Asym} (B) / Q$ of the sample \#1 at different heater powers $Q$ at $T=8$~K.
}
\label{Qdep}
\end{figure}

Figure~\ref{Qdep} shows the field dependence of $\Delta T_y^{Asym} (B) / Q$ at different heater powers $Q$ at $T=8$~K.
As shown in Fig.~\ref{Qdep}, $\Delta T_y^{Asym} (B) / Q$ converges to a straight line as $Q$ increases, indicating the good linearity  of $\Delta T_y^{Asym} (B)$ with respect to $Q$.
This linearity demonstrates that $\Delta T_y^{Asym} (B)$ is an intrinsic thermal Hall signal.

\begin{figure}[htbp]
\centering
\includegraphics[width= \linewidth]{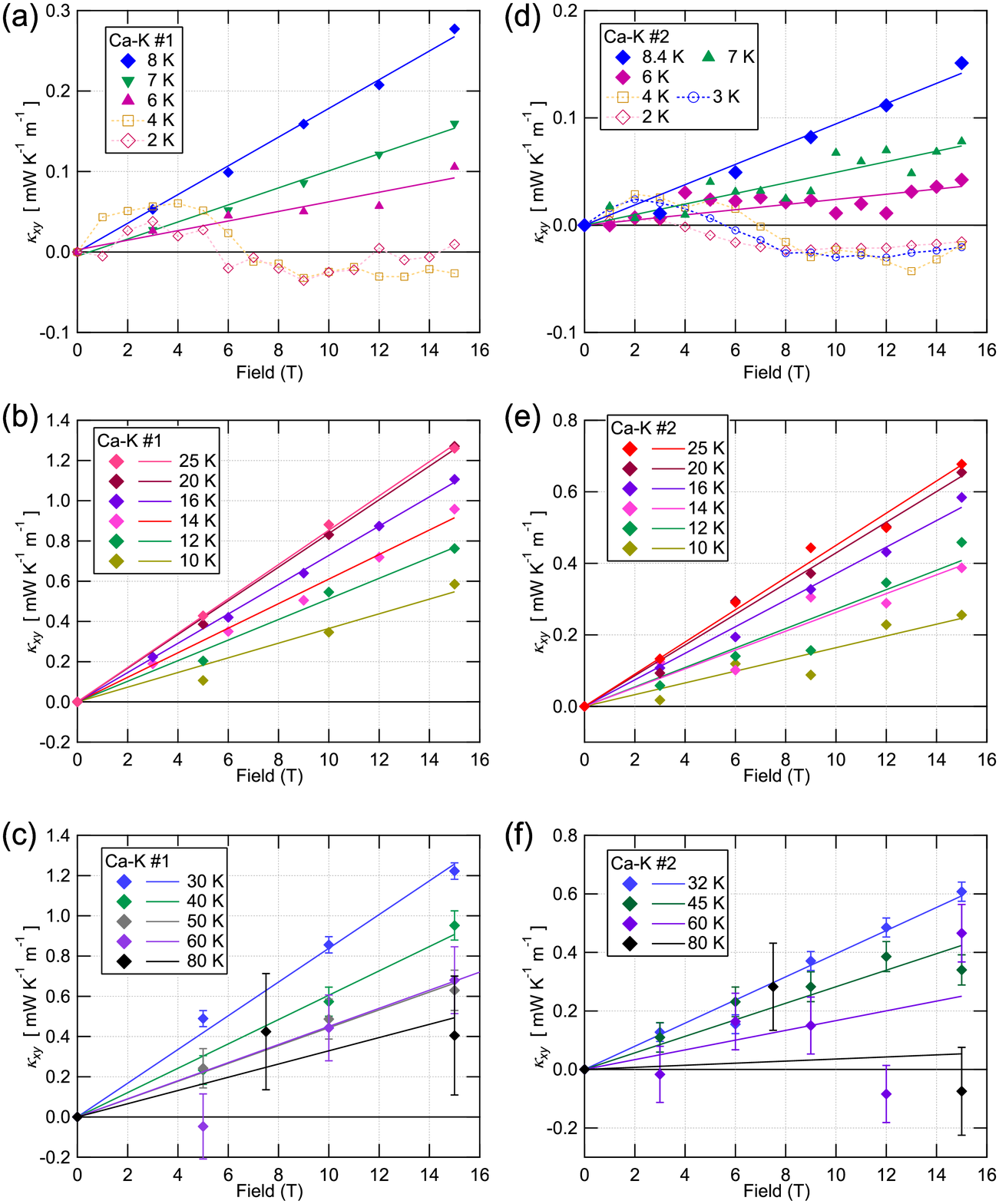}
\caption{
All field dependence data of $\kappa_{xy} (B)$ of sample \#1 (a-c) and sample \#2 (d-f).
The solid lines represent the linear fits of the data.
}
\label{All_kxy}
\end{figure}

Figure~\ref{All_kxy} shows all the field dependence data of $\kappa_{xy}$ of sample \#1 (a-c) and \#2 (d-f).
The data above $\sim 30$ K has a larger error in estimating $\kappa_{xy}$ (Fig.~\ref{All_kxy}(c) and (f)) because of the smaller signal and the smaller sensitivity of the Cernox thermometers at higher temperatures.
The error bars are smaller than the symbol size below 30 K.

\clearpage
\vspace{1em}
\vspace{1em}
\noindent{\bf S4.\ \ \  Calculation of spin thermal Hall conductivity by the Schwinger-boson mean-field theory}
\vspace{1em}

In order to study the thermal Hall conduction carried by spins in the quantum antiferromagnet materials, one may apply the linear response theory for the model

\begin{align}
 \mathcal{H}  = \frac{1}{2} \sum_{\langle i,j \rangle} \left( J\,\v S_i \cdot \v S_j + D_{ij}\, \v S_i\times \v S_j \cdot \hat{z} \right) - g \mu_B \sum_i \v B \cdot \v S_i,
	\label{eq:spin_H}
\end{align}
by evaluating the energy current in the spin language and then take the mean-field approximation of the slave particle\,(parton). 
Here, $D_{ij}$ denotes the Dzyaloshinskii-Moriya interaction, $g$ the $g$-factor, $\mu_B$ Bohr magneton, $\v B$ the magnetic field. 
However, as shown in Ref.\,\cite{Hyunyong2015}, one can start from the mean-field Hamiltonian in the parton representation first, which is much simpler, and obtain the exactly the same energy current operator and the linear response theory. Throughout this paper, we consider the Schwinger boson representation, i.e. $\v S_i = (1/2) \sum_{\alpha,\beta=\uparrow,\downarrow} b_{i\alpha}^{\dagger} {\bm \sigma}_{\alpha \beta} b_{i\beta}$ where ${\bm \sigma}$ stands for the Pauli matrices.

In the Schwinger-boson mean field theory, it is customary to introduce two kinds of mean-field parameters $\langle b_{i\sigma}b_{j\sigma'} \rangle$ and $\langle b_{i\sigma}^{\dagger} b_{j\sigma}\rangle$ to obtain the ground state self-consistently. According to the recent thorough theoretical investigation \cite{WangVishwanath2006}, four different classes are found to be possible on the Kagome lattice to describe the quantum disordered ground state. Among them, the so-called zero-flux and the $\pi$-flux classes are most often adopted to study the Kagome spin liquid (e.g. \cite{Sachdev1992}). The zero-flux state exhibits the $q=0$ magnetic order by condensing the spinons at low temperature, while the $\pi$-flux state comes to represent the $\sqrt{3} \times \sqrt{3}$ order. It is commonly known that the $q=0$ configuration takes energetic advantage from the DM interaction. Also, the recent numerical study \cite{Messio2017} suggests the $q=0$ ordered state as a possible ground state of Ca kapellasite. Hence, we employed the zero-flux ansatz where both $\langle b_{i\sigma}b_{j\sigma'} \rangle$ and $\langle b_{i\sigma}^{\dagger} b_{j\sigma}\rangle$ are involved. (On the other hand, the $\pi$-flux state requires only $\langle b_{i\sigma}b_{j\sigma'} \rangle$ term.) Besides, the $\langle b_{i\sigma}b_{j\sigma'} \rangle$ term stands for the creation of the singlet of neighboring spins and therefore it is reasonable to expect it to be less effective at a rather high temperature ($k_B T > 0.1 J$) we are considering. In addition, finding a solution of the self-consistent equations is greatly simplified by ignoring $\langle b_{i\sigma}b_{j\sigma'} \rangle$. For these reasons, we omitted the $\langle b_{i\sigma}b_{j\sigma'} \rangle$ term in the Ansatz.

Let us begin by decoupling the above Hamiltonian in terms of a mean-field ansatz $\chi_{ij,\sigma} \equiv \langle b^{\dagger}_{i\sigma} b_{j\sigma}\rangle$ which reads

\begin{align}
	\mathcal{H}^{\rm SBMF} = \sum_{i,\sigma} (\lambda - \sigma B) b_{i\sigma}^{\dagger} b_{i\sigma}
  + \sum_{\langle i,j \rangle, \sigma} \left( t^{\sigma}_{ij}
  b_{i\sigma}^{\dagger}b_{j\sigma} + h.c. \right),
\end{align}
where $t^{\sigma}_{ij} = J \chi_{ji}^{\sigma} + J' e^{-i\sigma\phi_{ij}} \chi_{ji}^{-\sigma}$, $J' =\sqrt{J^2 + D^2}$, and $\tan \phi_{ij} = D_{ij}/J$. Lagrange multiplier $\lambda$ is introduced to impose the constraint $2S=1$, and will be determined self-consistently. The sign of $D_{ij}$ is assumed to be positive if $i \rightarrow j$ is clockwise direction from the center of each triangle plaquette in the kagom\'e lattice, and we define $D_{ij} = - D_{ji} = D$.
In addition, we assume the effective hopping $t_{ij}^{\sigma}$ is independent on the bond, i.e., $\chi_{ij,\sigma} = \chi_{\sigma}$ and $t_{ij}^{\sigma}=t_{\sigma}$, throughout this article. By taking the Fourier transform of the above model into the momentum space, one obtains

\begin{align}
	& \mathcal{H}^{\rm SBMF} = \sum_{\bm{k},\sigma}\Psi_{\bm{k}\sigma}^{\dagger} \mathcal{H}_{\bm{k}\sigma}^{\rm SBMF} \Psi_{\bm{k}\sigma}, \nn
	& \mathcal{H}_{\bm{k}\sigma}^{\rm SBMF} =  (\lambda  - \sigma B) I_3 + \begin{pmatrix}
    	0 & t_{\sigma} \cos k_1 & t_{\sigma}^*\cos k_3\\
	    t_{\sigma}^*\cos k_1  & 0 & t_{\sigma} \cos k_2 \\
     	t_{\sigma}\cos k_3 &t_{\sigma}^*\cos k_2 & 0
  	\end{pmatrix}, 
  \label{eq:sb_H}
\end{align}
where $k_x=\bm{k}\cdot\hat{e}_x$, $\hat{e}_x$ are the three unit vectors specifying three sub-lattices\,($x=\alpha,\beta,\gamma$) of the kagom\'e lattice, and $\Psi_{\bm{k}\sigma}^T = [ b_{\bm{k}\sigma}^\alpha \,\,b_{\bm{k}\sigma}^\beta \,\,b_{\bm{k}\sigma}^\gamma ] $. Thermal Hall conductivity for the noninteracting bosonic Hamiltonian has been formulated in Ref.\,\onlinecite{Matsumoto2014}, which is also applicable to Eq.\,\eqref{eq:sb_H} resulting in

\begin{align}
	  \kappa^{\rm SBMF}_{xy} = - \frac{k_B^2 T}{\hbar N_t}
  \sum_{\bm{k}, n, \sigma} \left[
    c_2\left( n_B\left(\frac{E_{n\v k \sigma}}{k_B T}\right) \right) - \frac{\pi^2}{3}
  \right] \Omega_{n\v k \sigma},
  \label{eq:kappa_xy}
\end{align}
where $k_B$ is the Boltzmann constant, $T$ the temperature, $N_t$ the system size\,(or the number of unit-cell), $n_B(x) = (e^{x}-1)^{-1}$ the Bose-Einstein distribution function, and $n$ denotes the $n$-th energy band\,($E_{n\bm{k}\sigma}$) obtained by diagonalizing Eq.\,\eqref{eq:sb_H}. Also, $\Omega_{n \v k \sigma} = i\langle\partial_{k_x}  u_{n\v k \sigma} | \partial_{k_y} u_{n\v k \sigma}\rangle + c.c.$ is the Berry curvature of each band, the function $c_2(x) =(1+x)\left( \ln \frac{1+x}{x} \right)^2 - \left(\ln x\right)^2 - 2 {\rm Li}_2(-x)$, and ${\rm Li}_2$ is the polylogarithm function\,\cite{Matsumoto2014}. Now, our task is to find $\chi_{\sigma}$ and $\lambda$ by solving the self-consistent equations: $\sum_{\sigma} \langle b_{i\sigma}^{\dagger} b_{i\sigma} \rangle = 2S$ and $\langle b_{i\sigma}^{\dagger} b_{j\sigma}\rangle = \chi_{\sigma}$. One can easily verify that, with the assumption of translational symmetric ansatz, the first self-consistent equation becomes

\begin{align}
	\frac{1}{6N_t} \sum_{n,\v k,\sigma} n_B\left(\frac{E_{n\v k \sigma}}{k_B T}\right) = S,
\end{align}
and we used the identity $(1/N_t)\sum_i e^{i \v k \cdot \v r_i} = \delta_{\v k = 0}$. Also, the second-self consistent equation can be recast as follows:

\begin{align}
	\chi_{\sigma} & = \langle b_{i\sigma}^{\dagger} b_{j\sigma} \rangle \nn &= \frac{1}{6 N_t} \sum_i \langle (b_{i+\hat{e}_2 \sigma}^{\gamma\dagger} + b_{i-\hat{e}_2 \sigma}^{\gamma \dagger}) b_{i \sigma}^{\beta} + (b_{i + \hat{e}_1 \sigma}^{\alpha \dagger} + b_{i -\hat{e}_1 \sigma}^{\alpha \dagger}) b_{i + \hat{e}_2 \sigma}^{\gamma} + (b_{i \sigma}^{\beta \dagger} + b_{i-2\hat{e}_1 \sigma}^{\beta \dagger}) b_{i-\hat{e}_1 \sigma}^{\alpha} \rangle \nn
	& = \frac{1}{3N_t^2}\sum_{\v k, \v q, i}\left( \cos k_2 \langle b_{\v k \sigma}^{\gamma \dagger} b_{\v q \sigma}^{\beta} \rangle + \cos k_3 \langle b_{\v k \sigma}^{\alpha \dagger} b_{\v q \sigma}^{\gamma} \rangle + \cos k_1 \langle b_{\v k \sigma}^{\beta \dagger} b_{\v q \sigma}^{\alpha} \rangle \right) e^{i(\v k - \v q) \cdot \v r_i}\nn
	& = \frac{1}{3N_t}\sum_{\v k}\left( \cos k_2 \langle b_{\v k \sigma}^{\gamma \dagger} b_{\v k \sigma}^{\beta} \rangle + \cos k_3 \langle b_{\v k \sigma}^{\alpha \dagger} b_{\v k \sigma}^{\gamma} \rangle + \cos k_1 \langle b_{\v k \sigma}^{\beta \dagger} b_{\v k \sigma}^{\alpha} \rangle \right),
\end{align}
and therefore

\begin{align}
	{\rm Re} \chi_{\sigma} & = \frac{1}{2}(\chi_{\sigma} + \chi_{\sigma}^*) 
	= \frac{1}{6 N_t} \sum_{\v k} \left\langle \begin{pmatrix}
		b_{\v k \sigma}^{\alpha \dagger} & b_{\v k \sigma}^{\beta \dagger} & b_{\v k \sigma}^{\gamma \dagger}
	\end{pmatrix}	
	\begin{pmatrix}
    	0 & \cos k_1 & \cos k_3\\
	    \cos k_1  & 0 & \cos k_2 \\
     	\cos k_3 & \cos k_2 & 0
  	\end{pmatrix}
	\begin{pmatrix}
		b_{\v k \sigma}^{\alpha} \\ b_{\v k \sigma}^{\beta} \\ b_{\v k \sigma}^{\gamma}
	\end{pmatrix}
	\right\rangle, \nn
	& = \frac{1}{6 N_t} \sum_{\v k, n} f_{n \v k \sigma}(t_\sigma = 1) \langle b_{n \v k \sigma}^{\dagger} b_{n \v k \sigma}\rangle
	= \frac{1}{6 N_t} \sum_{\v k, n} f_{n \v k \sigma}(t_\sigma = 1) n_B\left( \frac{E_{n \v k \sigma}}{k_B T} \right),
\end{align}
and 

\begin{align}
	{\rm Im} \chi_{\sigma} & = \frac{1}{2i}(\chi_{\sigma} - \chi_{\sigma}^*) 
	= \frac{1}{6 N_t} \sum_{\v k} \left\langle \begin{pmatrix}
		b_{\v k \sigma}^{\alpha \dagger} & b_{\v k \sigma}^{\beta \dagger} & b_{\v k \sigma}^{\gamma \dagger}
	\end{pmatrix}	
	\begin{pmatrix}
    	0 & \cos k_1 & -\cos k_3\\
	    -\cos k_1  & 0 & \cos k_2 \\
     	\cos k_3 & -\cos k_2 & 0
  	\end{pmatrix}
	\begin{pmatrix}
		b_{\v k \sigma}^{\alpha} \\ b_{\v k \sigma}^{\beta} \\ b_{\v k \sigma}^{\gamma}
	\end{pmatrix}
	\right\rangle, \nn
	& = \frac{1}{6 N_t i} \sum_{\v k, n} f_{n \v k \sigma}(t_\sigma = i) \langle b_{n \v k \sigma}^{\dagger} b_{n \v k \sigma}\rangle
	= \frac{1}{6 N_t i } \sum_{\v k, n} f_{n \v k \sigma}(t_\sigma = i) n_B\left( \frac{E_{n \v k \sigma}}{k_B T} \right),
\end{align}
where $f_{n \v k \sigma}(t_\sigma = 1)$ is the eigen-function of off-diagonal part of Hamiltonian in Eq.\,\eqref{eq:sb_H} with $t_\sigma = 1$.

\begin{figure}[htbp]
\centering
\includegraphics[width=1.0\linewidth]{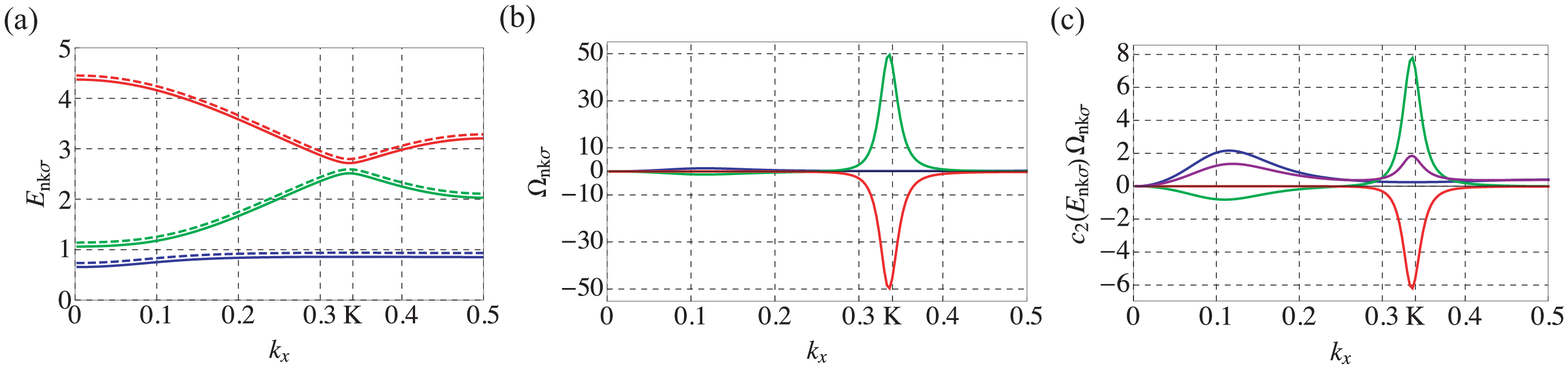}
\caption{ (a) Energy bands, (b) Berry curvatures, and (c) Berry curvature multiplied by $c_2$ function\,(see text) as a function of $k_x$ in unit of $\pi$ at $k_y=0$, $k_BT=J$, $D=0.1J$ and $g \mu_B \v B/J = 0.01 \hat{z}$. 
The solid\,(dashed) lines are the bands with up-\,(down-)spin bosons. The red\,(green/blue) lines are for the upper\,(middle/lower) band.
The purple line in (c) shows the sum of the all bands.
For clarity, only up-spin bands are shown in (b) and (c). }
\label{fig:band_berry}
\end{figure}

\begin{figure}[htbp]
\centering
\includegraphics[width=0.6\linewidth]{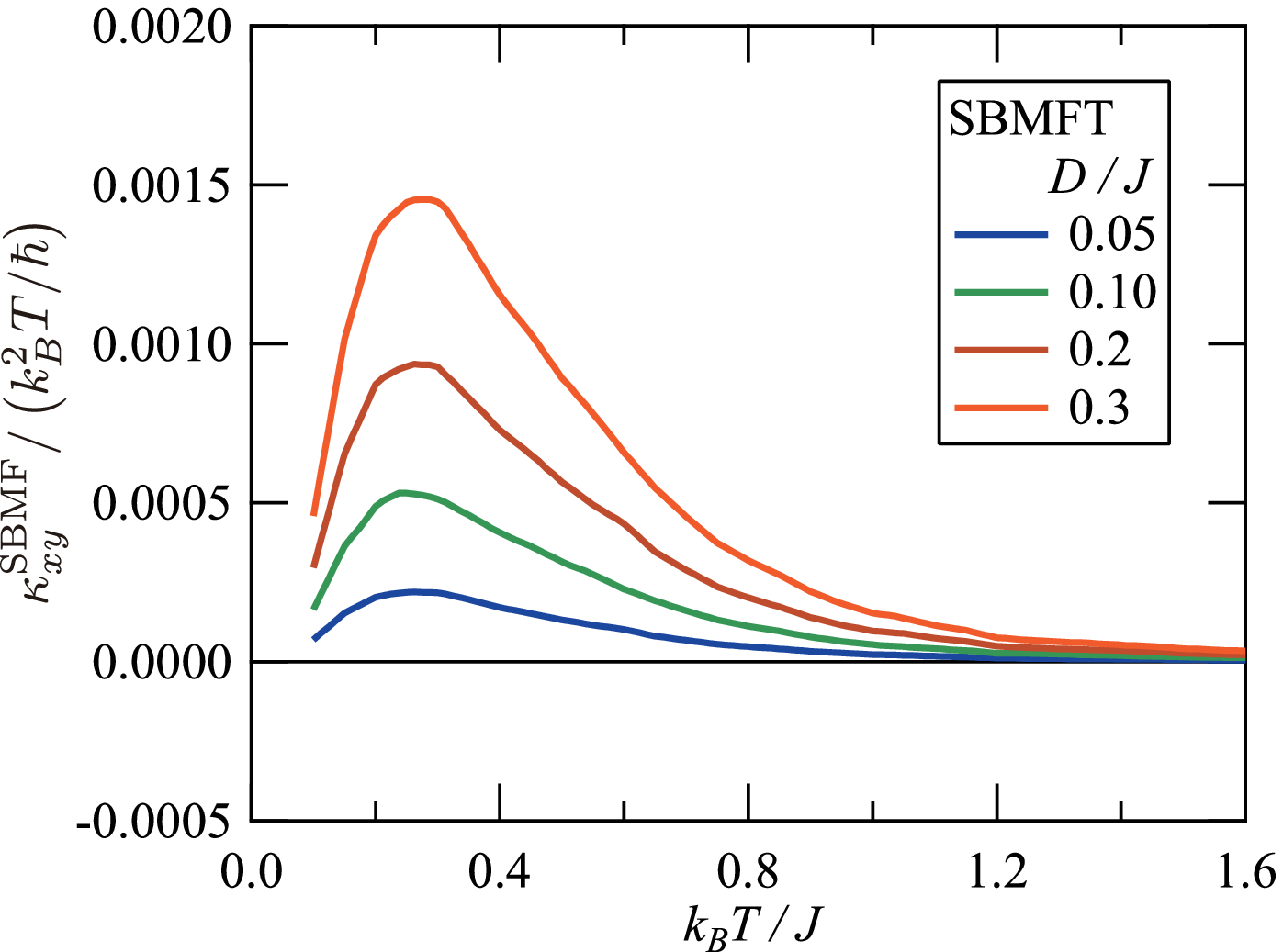}
\caption{
SBMF calculations of $\kappa_{xy}^{\rm SBMF}$ at $D/J = $ 0.05, 0.1, 0.2, and 0.3 at $g \mu_B \v B/J = 0.01 \hat{z}$.
}
\label{fig:SBcalc}
\end{figure}

We solve the above equations numerically and determine the effective hopping integral $t_{\sigma}$ as functions of the temperature and magnetic field. As an example, we present the energy band and Berry curvature at $k_B T=J$ and $g \mu_B \v B/J = 0.01 \hat{z}$ in Fig.\,\ref{fig:band_berry}, where the obtained $t_{\uparrow}=1.1246+0.34668i$, $t_{\downarrow} =1.137+0.35666i$ and $\lambda = 2.2$.
Due to the finite magnetic field, the bands with up-spin\,(solid lines in Fig.\,\ref{fig:band_berry}\,(a)) and down-spin\,(dashed lines in Fig.\,\ref{fig:band_berry}\,(a)) are shifted in an opposite way, and therefore a small gap between two bands occurs as one can expect from Eq.\,\eqref{eq:sb_H}\,[Fig.\,\ref{fig:band_berry}\,(a)]. In Fig.\,\ref{fig:band_berry}\,(b) and (c), the red\,(green/blue) line is the Berry curvature and  the Berry curvature multiplied by $c_2$ function of the upper\,(middle/lower) band of up-spin Schwinger boson, respectively. Down-spin SB bands have exactly opposite Berry curvatures at $|\v B| = 0$ and a similar one throughout $|\v B| > 0$. Using $t_{\sigma}$ at each temperature and magnetic field, one can evaluate the thermal Hall conductivity by performing the integration in Eq.\,\eqref{eq:kappa_xy}, and the results for several $D$ and $g \mu_B \v B/J = 0.01 \hat{z}$ are shown in Fig.\,\ref{fig:SBcalc} as a function of temperature. As shown in the main text, $\kappa_{xy}^{\rm SBMF}$ shows a peak around $k_BT/J \sim 1/3$. Note that $\kappa_{xy}^{\rm SBMF}$ depends linearly on the Dzyaloshinskii-Moriya interaction while the peak position does not.

\vspace{1em}

\clearpage
\vspace{1em}
\vspace{1em}
\noindent{\bf S5.\ \ \  Values of $J$ and $D/J$ of Ca kapellasite and volborhite obtained by different methods.}
\vspace{1em}

We summarize the data of $J$ and $D$ obtained by our study and by other methods in Table S1. 
In the row of ``$\chi$, $\Delta g/g$", we take $J$ value from the high-temperature fitting of the magnetic susceptibility data and estimate $|D/J|$ by $(g_c-2)/2$.
In the first principles result of volborthite, we estimate $J_{eff}=(J+J_1+J')/3$
(see Ref.~\cite{Janson2016} for the definition of $J$, $J_1$, and $J'$) because of the anisotropic spin Hamiltonian.

\begin{table}[h]
 \caption{Values of $J$ and $|D/J|$ of Ca kapellasite and volborhite obtained by different methods.}
 \scalebox{1.5}{
 \begin{tabular}{|c|c|c|c|} \hline
	Materials & Methods & $J/k_B$ & $|D/J|$ \\ \hline
		& $\chi$, $\Delta g/g$~\cite{HYoshida2017}	& 52.2	& 0.07 \\ \cline{2-4}
	Ca kapellasite	& SBMFT fitting to $\kappa_{xy}$	& 66	& 0.06--0.12 \\ \cline{2-4}
		& First principles~\cite{JeschkePrv}		& 64	& N/A	\\ \hline
		& $\chi$~\cite{Hiroi2001}, $\Delta g/g$~\cite{Ishikawa2015}	& 84	& 0.14 \\ \cline{2-4}
	Volborthite	& SBMFT fitting to $\kappa_{xy}$	& 60	& 0.07 \\ \cline{2-4}
		& First principles~\cite{Janson2016}		& 25	& 0.07	\\ \hline
 \end{tabular}
 }
\end{table}

\end{document}